\documentclass[11pt]{amsart}

\usepackage{graphicx,amsmath,amssymb,fullpage,stackrel}
\usepackage{subeqnarray}

\title{AN ENZYMATIC HORMESIS BOX}
\author{Michael Grinfeld}
\address{Department of Mathematics and
    Statistics, University of Strathclyde, 26 Richmond Street, Glasgow
    G1 1XH, UK}
 \email{m.grinfeld@strath.ac.uk}

\begin{document}

\begin{abstract}
\noindent We present a simple enzymatic system that is capable of a
biphasic response under competitive inhibition. This is arguably the
simplest system that can be said to be hormetic.\\

\end{abstract}

\maketitle

\section{Introduction}

Though 21st century biology is excellent in collecting data, and very
good at translating these data into therapeutical interventions, it is
less good at identifying principles of biological organisation.

It has long been suggested (see for example \cite{Forbes}) that
hormesis is a principle of biological organisation. Whether it is or
is not, which is a highly controversial matter as we discuss below, it
would be useful to present simple mechanisms at any level of
biological organisation that allow for hormesis. The present paper
takes no position in the heated debate about hormesis and restricts
itself to suggesting an enzymatic ``hormesis box'', which might be of
independent interest to students of enzymology.

\section{An introduction to hormesis}

To quote Wikipedia \cite{wiki}, {\bf Hormesis} ``is any process in a
cell or organism that exhibits a biphasic response to exposure to
increasing amounts of a substance or condition.''

We have chosen this definition as it is ``ethically neutral'': it
makes no claim about the substance or condition being beneficial or
noxious (injurious).

\begin{figure}[htp]
  \centerline{\includegraphics[width=.6\textwidth]{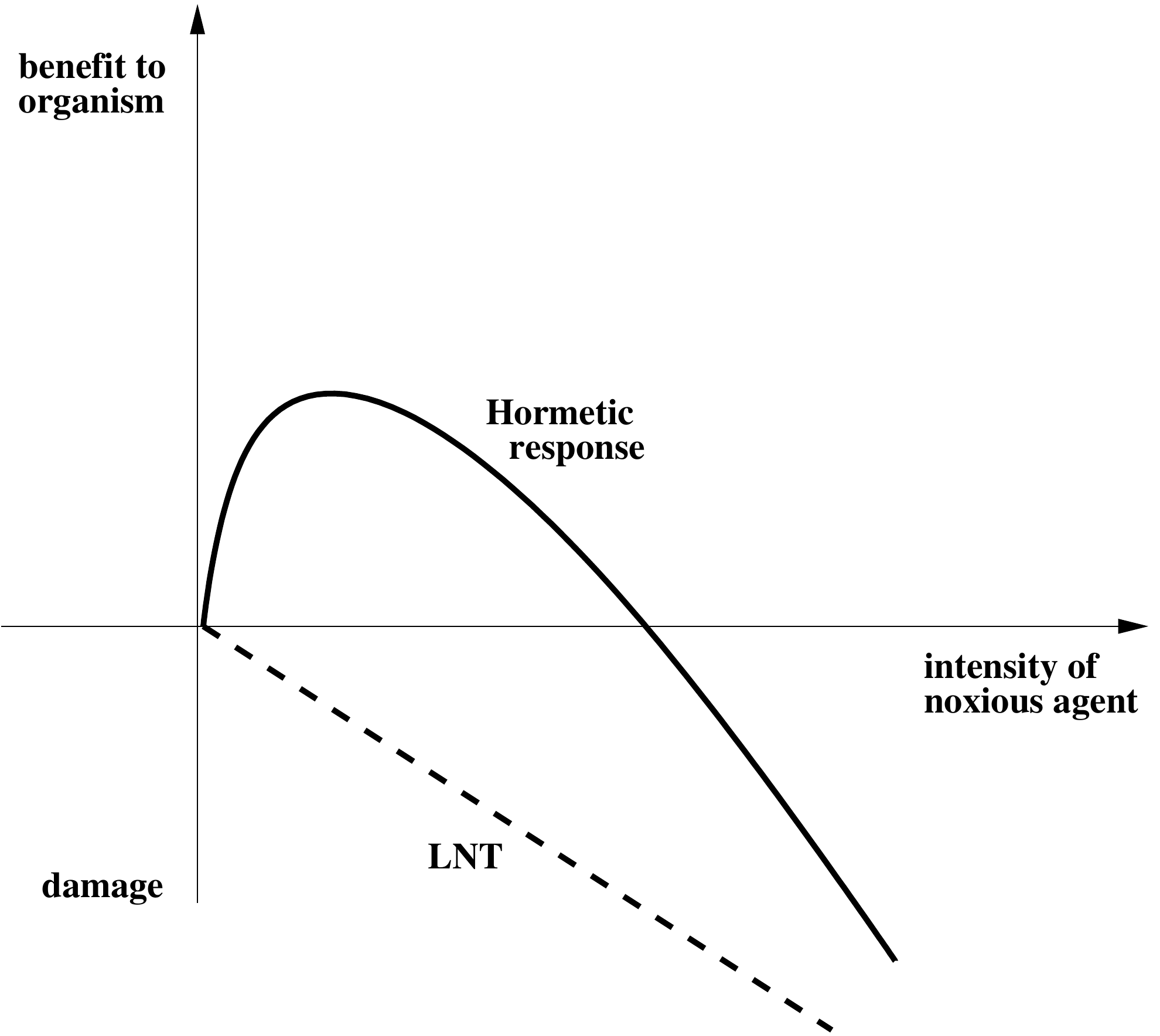}}
  \caption{A hormetic dose response curve.} \label{fig:1}.
\end{figure}

A typical hormesis dose-response curve which is not so ethically
neutral is shown in Figure~\ref{fig:1}. Here the agent is assumed to
be injurious and by a hormetic response in such a situation one
(contentiously) means a response in which an injurious agent in small
concentration confers benefits on the organism; we chose to label the
axes in this way to alert the reader to the crux of the controversy.

In Figure~\ref{fig:1} LNT stands for ``Linear-No-Threshold'', the
basis of much health and safety legislation.  If instead of
``noxious'' one puts ``beneficial'' (as in anti-cancer drugs), one
should reflect the hormetic and the LNT responses around the abscissa,
both, again, debatable statements

To see the fervour with which the merits and demerits of hormesis as a
general principle of organisation are discussed, consult, for example
\cite{Sacks}. In our opinion, the clearest (though partisan) analysis
of the difficulties with hormesis is in \cite{SF}[Ch. 3]; that chapter
is entitled ``Hormesis Harms: the emperor has no biochemistry
clothes'' .

Shrader-Frechette distinguishes three different assertions: [H], [HG]
and [HD], and says that [H] is trivially true, [HG] is demonstrably
false, and [HD] is ethically and scientifically questionable. Here 

\begin{itemize}
\item{{\bf [H]:}} (In a particular biological context) there exists an
endpoint (an observable) for which a noxious agent exhibits a
beneficial effect at low concentrations; 

\item{{\bf [HG]:}} [H] is generalizable across different biological
  contexts, endpoints measured, and classes of chemicals; 

\item{{\bf [HD]:}} [H] should be the default assumption in the risk-assessment 
[and hence in risk-regulation] processes.
\end{itemize}

We comment that methodologically it is not clear by what criteria [HG]
can be derived from instances of [H]: what does ``generalizable''
really mean? and that [HD] cannot follow from [HG], no matter how much
money it saves to a particular industry; LNT will always be much
better equipped to withstand legal challenges.

However, the impetus for the present work is Shrader-Frechette's
statement that [H] is ``trivially true''. While Shrader-Frechette
admits that [H] is well documented across plant, fungal and animal
kingdoms, there must arise the question why this should be the case.

Our tentative answer is that hormesis must hitch a ride on some very
general principle (or principles) of biological organisation and does
not make this principle (or these principles) fitness-reducing.

\section{A simple enzymatic model of hormesis}

We consider a system comprised of an enzyme $E$,
a substrate $S$, a product $P$  and an inhibitor $I$. By hormesis in
such a system we mean that the rate of production of $P$ should increase
on adding sufficiently small amounts of the inhibitor. Obviously,
arbitrarily complex mechanisms leading to hormesis can be devised,
involving, for example, additional transcription of genes encoding the
enzyme $E$. We aim here for maximal simplicity and ask what frequently
encountered mechanism can be superimposed on a single enzyme
confronted with a (competitive) inhibitor so that the system is
hormetic. 

We start with the indisputable fact that many proteins exist in a
dimeric form (there is even a special BTB ``born-to bind'' domain!)
\cite{BTB}. The intuition behind the mechanism that we are proposing
is as follows: in the absence of the inhibitor, most molecules of the
enzyme $E$ are to be found in the inert, dimer form $E_2$. In addition
to combining with the enzyme at its active site, the inhibitor also
causes the dissociation of the dimer, this making more active enzyme
available for the production of $P$.

Therefore the reactions we must consider are: 
\begin{subeqnarray}\label{scheme}
  \slabel{sc1}
  E+E   & \stackrel[k_2]{k_1}{\rightleftharpoons} & E_2,\\
  \slabel{sc2}
  E+I   & \stackrel[k_4]{k_3}{\rightleftharpoons} & EI,\\
  \slabel{sc3}  
     E+S   &\stackrel[k_6]{k_5}{\rightleftharpoons} & ES
     \stackrel{k_7}{\rightharpoonup} E+P,\\
   \slabel{sc4}  
     E_2+I & \stackrel[k_9]{k_8}{\rightleftharpoons} & E_2 I
     \stackrel{k_{10}}{\rightharpoonup}{} 2E+I.
\end{subeqnarray}

We assume that the substrate $S$ is in constant supply. We would like
to understand under what conditions on the the rate of production of
$P$, i.e. the quasi-steady state concentration of the complex $ES$
increases as we add a small amount of the inhibitor.   

Below we denote concentrations of species by square brackets enclosing
the symbol of the species. The system (\ref{scheme}) and the theory of
enzyme kinetics \cite{CB} implies that we need to find the expression
for the concentration of $ES$ from the following system of two
conservation laws for the total amount of the enzyme and the
inhibitor,
\begin{subeqnarray}\label{cl}
  \slabel{clE}
  [E_0] & = & [E]+2[E_2]+[EI]+[ES]+2[E_2I],\\
  \slabel{clI} [I_0] & = &[I] +[EI] +[E_2I],
\end{subeqnarray}
and the four equations we get by making quasi-steady state
assumptions for $ES$, $EI$ and $E_2I$:
\begin{subeqnarray}\label{qssa}
\slabel{e3}
  k_3[E][I] & = &k_4[EI],\\
  \slabel{e4}
  k_5[E][S] & = &(k_6+k_7)[ES],\\
  \slabel{e5}
k_8[E_2][I] & = & (k_9+k_{10})[E_2I],\\
\slabel{e6}
k_1[E]^2 & = & 2k_2[E_2]+2k_{10}[E_2I].
\end{subeqnarray}

Solving the system of equations (\ref{cl})--(\ref{qssa}) purely
symbolically is a formidable task as the scheme (\ref{scheme})
involves ten kinetic constants $k_1, \ldots, k_{10}$, and we also have
to take into account stoichiometric constraints. Of course, $4$ of the
kinetic constants can be absorbed and non-dimensionalisation will get
rid of one of the stoichiometric constraints. The remaining system
will still have eight parameters, and as the goal of the paper is not
an exhaustive analysis of (\ref{cl})--(\ref{qssa}), we make rather
crude simplifications in order to exhibit the possibility of hormesis in this
system. Intuitively, to get a hormetic response, we need both $k_1$
and $k_{10}$ to be ``large'', so that without inhibitor most enzyme is
in dimeric form $E_2$, and, once inhibitor is added, it causes enough
release of the active enzyme $E$ from the pool sequestered in the
dimer to raise the level of production of $P$. To exhibit the relation
between the strengths of the two processes that results in a hormetic
response, we simply put
\[
  k_2 = k_3 = \cdots = k_9 = 1,
\]
and take $[E_0]=1$, i.e. set the total concentration of
enzyme equal to one, and assume that the concentration of the
substrate $S$ is constant, and satisfies $[S] \ll [E_0]$; in our
computations below we use $[S]=10$.

Thus we have a system of six equations in the variables $[E]$, $[ES]$,
$[EI]$, $[E_2]$, $E_2I]$ and $[I]$ with symbolic parameters
$k_1, k_{10}$ and $[I_0]$. Since
\[
  \frac{d[P]}{dt}= k_7[ES],   
\]
we are really only interested in $[ES]$ as a function of these
parameters. Obtaining an explicit formula for $[ES]$ (satisfying
$0< [ES] < [E_0]=1$), is still nontrivial, but a univariate polynomial
satisfied by $[ES]$ with coefficients depending on $k_1, k_{10}$ and
$[I_0]$, $P([ES],[I_0],k_1,k_{10})$ can be easily obtained in MAPLE \cite{Maple}
using the \verb+Groebner+ package (it is denoted by \verb+poly+ in
the code in Appendix \ref{App}. For more information on Gr\"obner
bases that have proved to be very useful in solving polynomial
equations of enzyme kinetics, the reader is referred to \cite{CLO}.

\begin{figure}[htp]
\centerline{
  \includegraphics[width=.8\textwidth]{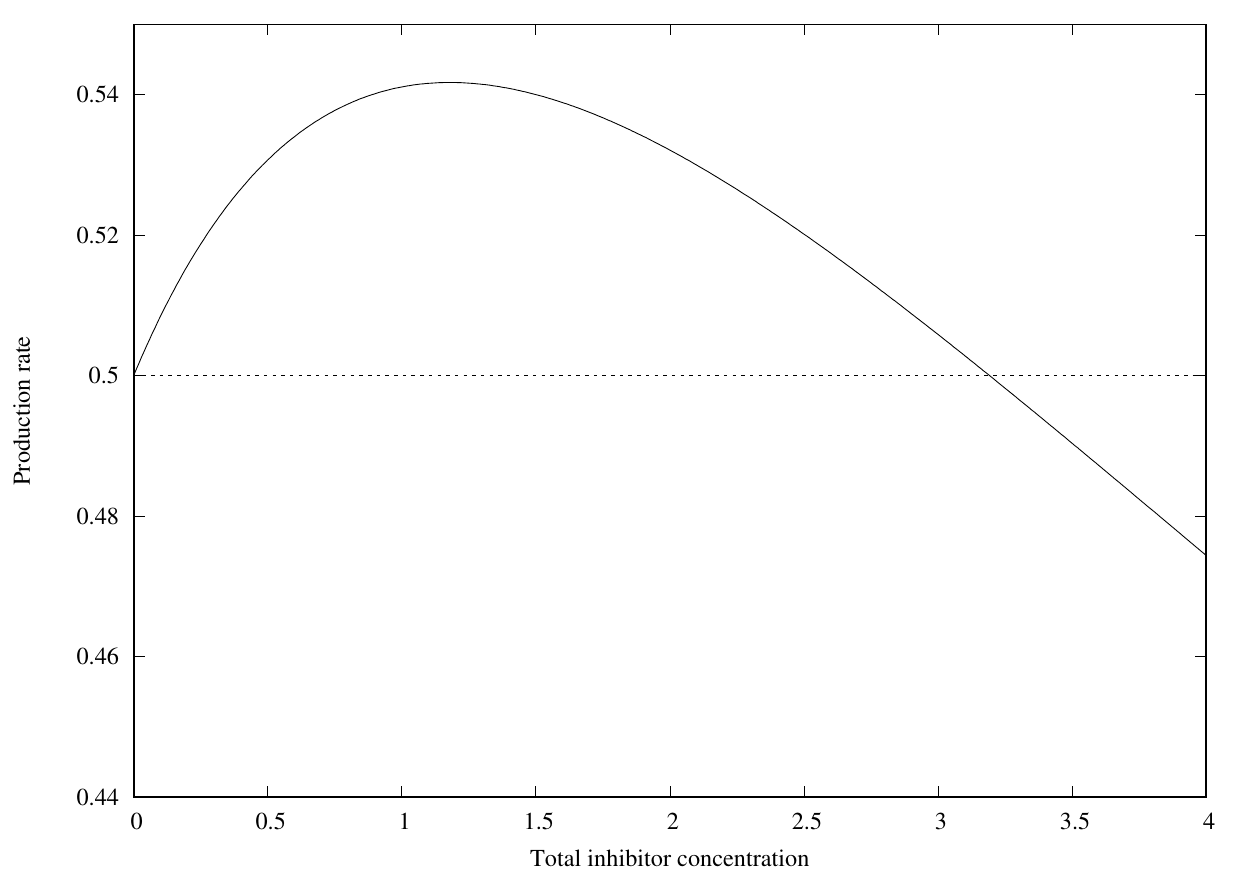}}
\caption{The hormetic response in (\ref{scheme}); see text for
  parameter values.} \label{fig:2} 
\end{figure}

Solving $P([ES],[I_0],k_1,k_{10})=0$, we have that the
rate of production of $P$ in the absence of the inhibitor, i.e. with
$[I_0]=0$, which we denote by $r_0$ is given by 
\[
  r_0= \frac{5(-3+\sqrt{9+k_1})}{k_1}.
\]
Now we use a regular perturbation expansion: we write
\[
  [ES]= r_0+ r_1 [I_0]+ O([I_0]^2).
\]
Hence hormesis is equivalent to $r_1>0$. Computing $r_1$ shows that
the condition for hormesis can be written in a  very elegant way:
\begin{equation}\label{k1crit}
  k_1 > k_{1,crit} := \frac{7 k_{10}^2+2k_{10}-5}{(k_{10}-1)^2}, \quad
  k_{10}>1.
\end{equation}
From (\ref{k1crit}) it is clear that hormesis is not possible for any
value of $k_{10}$ is $k_1\leq 7$. In the simulation below we choose
$k_1=40$, $k_{10}=50$ which clearly falls in the hormetic regime of
(\ref{k1crit}), and from $P([ES],[I_0],40,50)=0$ find the unique value
of $[ES]$ in $[0, 1]$ as a function of $[I_0]$. The results are
presented in Figure~\ref{fig:2}, and clearly show biphasic response.

\section{Discussion}

As observed in the Introduction, in Figure~\ref{fig:1} the ordinate is
in units of ``benefit'' to the organism as most emotionally charged
discussions of ethics and philosophy of hormesis are couched in these
terms.  We do not commit to interpret the (counterintuitive) increase
in the rate of production of $P$ in our model system as the
competitive inhibitor is added to a benefit, being entitled to do so
by our ethically neutral definition of hormesis.

In this short paper we presented a plausibly general mechanism by
which a hormetic biphasic response could be generated in a simple
enzymatic system; what made the biphasic response possible was
sequestration of most enzyme in a dimer and the inhibitor-regulated
release of the monomers from the dimer. It would be interesting to know
if systems where this mechanism is operational exist, and if they do,
how frequent are they.

The claim [HG] in ethically neutral terms states that biphasic effects
are the norm. If true, this is in fact as astonishing claim as, in our
context, it does not mention the inhibitor ever being encountered
before in the history of a species; it could be a chemical that had
been synthesized for the first time in history just before the
exposure experiment.

Many of R. Rosen's examples of anticipatory systems \cite{Rosen} can
be explained away by recourse to an evolutionary argument: the system
behaves in such and such a way (e.g. the product of the first in a
chain of enzymatic reactions leads to an activation of transcription
of the last enzyme in the chain) because such behaviour is ``fitter''
than its absence (as, in the above example, absence of such an
``anticipatory'' effect would lead to a buildup of an intermediate in
the chain of reactions), in which case the description of such a
system as ``anticipatory'' is redundant and has no explanatory power.
Hence an ethically neutral form of [HG], if true, would indicate the
existence of intrinsically anticipatory systems.

\appendix

\section{Maple code for computations used in this paper}
\label{App}

Here we show how to compute the polynomial in $[ES]$, \verb+poly+ and
the relation between $k_1$ and $k_{10}$ (\verb+k1crit+ below).  Below
\verb+ES+ corresponds to $[ES]$m etc.

\begin{verbatim}
with(Groebner):

# set E0 to 1 and S to 10
# we use I1 instead of I as I is reserved I Maple

E0:= 1: S := 10:

# conservation laws

e1:= E0-E-2*E2-EI-ES-2*E2I:
e2 := I0-EI-E2I-I1:

# set all constants equal to 1 apart from k1 and k10

k2:=1: k3:=1: k4:=1: k5:= 1: k6:=1: k7:=1: k8:= 1: k9:=1:

# QSSAs:

e3:= k3*E*I1-k4*EI:

e4 := -k7*ES-k6*ES+k5*E*S:

e5 := k8*E2*I1-k9*E2I-k10*E2I:

e6 := -k1*E^2+k2*2*E2+2*k10*E2I: # more terms here but all zero by QSSA. 

F := [e1,e2,e3,e4,e5,e6];

poly:= Basis(F,plex(I1,E,E2,EI,E2I,ES))[1]: 

# poly is a 5th order polynomial in ES with coefficients that depend on
# I0, k1 and k10.  You need the solution of poly that lies between
# zero and E0.

ab := factor(subs(I0=0,poly)):

r0:= solve(op(1,ab),ES)[1]:

# r0 is the value of ES when I0=0

ac := subs(ES=r0+r1*I0,aa):

# this is the regular perturbation expansion

ad:= coeff(ac,I0,1):

r1 := solve(ad,r1):

k1crit:= solve(r1,k1);

# gives the minimal value k1 must have as a function of k10 to give 
# hormetic response

\end{verbatim}


\begin{thebibliography}{99}
\bibitem{CB} A. Cornish--Bowden, {\em Fundamentals of Enzyme
    Kinetics}, John Wiley and Sons, Weinheim 2013.
\bibitem{CLO} D. Cox, J. Little, and D. O'Shea, {\em Ideals, Varieties,
    and Algorithms: an Introduction to Computational Algebraic
    Geometry and Commutative Algebra}, Springer, New York 2013. 
 \bibitem{Forbes} V. E. Forbes, Is hormesis an evolutionary
   expectation? Funct. Ecology {\bf 14} (2000), 14--24.    
\bibitem{Maple} Maplesoft, a division of Waterloo Maple Inc., {\em
    Maple}, Waterloo, Ontario 2019.
\bibitem{BTB} R. Perez-Torrado, D. Yamada, and P.-A. Defossez, Born to
  bind: the BTB protein-protein interaction domain, Bioessays {\bf 28}
  (2006), 1194-1202.
\bibitem{Rosen} R. Rosen, {\em Anticipatory Systems}, Springer, New
  York 2012.
\bibitem{Sacks} B. Sacks, G. Meyerson, and J. A. Siegel, Epidemiology
  without biology: false paradigms, unfounded assumptions, and
  specious statistics in radiation science, Biol. Theory {\bf 11}
  (2016), 69--101. 
\bibitem{SF} K. Shrader-Frechette, {\em  Tainted}, Oxford University
  Press, Oxford 2014. 
\bibitem{wiki} \verb+https://en.wikipedia.org/wiki/Hormesis+
\end{thebibliography}
\end{document}